\documentclass[runningheads]{llncs}
\usepackage{graphicx}
\usepackage{microtype}
\usepackage{soul}
\usepackage{times}
\usepackage{epsfig}
\usepackage{amsmath}
\usepackage{amssymb}
\usepackage{multirow} 
\usepackage{cancel}
\usepackage{rotating}
\usepackage{latexsym}
\usepackage{adjustbox}
\usepackage[thinlines]{easytable}
\usepackage{subcaption}
\usepackage{booktabs}       
\usepackage{mfirstuc}
\usepackage{makecell}
\usepackage{url}

\usepackage[pagebackref=true,breaklinks=true,colorlinks,bookmarks=false]{hyperref}
\newtheorem{assumption}{Assumption}

\usepackage{color,colortbl,array,xspace}
\usepackage[disable]{todonotes}
\def\yinzhi#1{\todo[backgroundcolor=yellow]{ycao: #1}\xspace}

\begin{document}
\title{Defending Medical Image Diagnostics against Privacy Attacks using Generative Methods: Application to Retinal Diagnostics
}
\titlerunning{Defending against Privacy Attacks of Image Diagnostics}
\author{William Paul\inst{1}
\and
Yinzhi Cao\inst{2}
\and
Miaomiao Zhang\inst{3}
\and
Phil Burlina\inst{1,2,4}
}
\authorrunning{Paul, Cao, Zhang, and Burlina}

\institute{ Johns Hopkins University, Applied Physics Laboratory, Laurel, MD 
\\
\and
 Johns Hopkins University, Dept. of Computer Science \\\and
University of Virginia, Dept. of Electrical Engineering
 \and
 Johns Hopkins University, Malone Center for Engineering in Healthcare
\email{william.paul@jhuapl.edu}
}
\maketitle              
\begin{abstract}

Machine learning (ML) models used in medical imaging diagnostics can be vulnerable to a variety of privacy attacks, including  {\em membership inference attacks}, that lead to violations of  regulations governing the use of medical data and threaten to compromise their effective deployment in the clinic. In contrast to most recent work in privacy-aware ML that has been focused on model alteration and post-processing steps, we propose here a novel and complementary scheme that enhances the security of medical data by controlling the data sharing process. We develop and evaluate a privacy defense protocol based on using a generative adversarial network (GAN) that allows a {\em medical data sourcer} (e.g. a hospital) to provide an external agent ({\em a modeler}) a proxy dataset synthesized from the original images, so that the resulting diagnostic systems made available to {\em model consumers} is rendered resilient to privacy {\em attackers}. 
We validate the proposed method on retinal diagnostics AI used for diabetic retinopathy that bears the risk of possibly leaking private information. To incorporate concerns of both privacy advocates and modelers, we introduce a metric to evaluate privacy and utility performance in combination, and demonstrate, using these novel and classical metrics, that our approach, by itself or in conjunction with other defenses, provides state of the art (SOTA) performance for defending against privacy attacks. 

\end{abstract}

\section{Introduction}
There has been a recent proliferation of artificial intelligence (AI) and machine learning (ML)  applications being developed and proposed for deployment in various tasks ranging from vision \cite{resnet,dense,imageRec} to natural language processing and speech~\cite{speech,radford2019language,rogers2021primer,vaswani2017attention}. However, ensuring guarantees of privacy for the data used for training those applications and for medical and retinal AI diagnostics~\cite{burlina2011automatic,burlina2017automated,esteva2017dermatologist,gulshan2016development,pekala2019deep,ting2019artificial,ting2019deep,topol2019high} is shaping up as an open impediment to deployment. For the purposes of this work, we focus on ensuring privacy when a trained classification model (classifier) created by a {\em modeler} is accessible to an attacker, allowing them to acquire information about individuals whose data was used in the training process. The attacker may use the model to infer private attributes (e.g. age or co-morbidities) about a specific person, in what is called an {\em attribute inference attack} \cite{fredrikson2014privacy}, or to determine if an individual's data was used for training, in what is termed a {\em membership inference attack}~\cite{NDSS,earliest,globalloss},
which is the focus of this study. Notably, we focus in this work on the case of the data being stored in a central location under the provenance of a trusted agent called a {\em medical data sourcer}, acting under an institutional review board that provides for an individual's privacy. Although we focus on having a single data sourcer for this work, we believe this work can be extended to the federated scenario \cite{kaissis2020secure,vizitiu2019towards} where there are multiple sources of data. 

This work focuses on how access to the data can be controlled by the medical data sourcer, namely creating new data points that should not contain the true identities of the original individuals that can safely be passed to modelers. This work evaluates the effectiveness of synthetic data for privacy, on retinal imagery collected for the task of diabetic retinopathy, alongside other defenses that affect the classifier directly. Due to the degree of data access being controlled, in terms of how many private data points are shared, there is a question of how to capture the typical trade-off between the privacy conferred by the model as well as the performance. Consequently this work introduces an additional metric that attempts to capture the trade off in a single measure, allowing for negotiation about the level of access between the data sourcer and modeler.

The unique and salient contributions of this work include: 

\begin{enumerate}
    \item We develop a novel strategy for privacy defense based on rejecting potentially vulnerable samples from generative models, which only depends on the source of the data, and requires no change in procedure in training. We believe that this approach can be used more broadly for other image-based medical diagnostics and other image classification tasks beyond healthcare.
    \item We propose a novel metric, called P1-score, to measure the trade-off between utility and privacy, so that privacy advocates and modelers can view both concerns together in contrast to most methods which only look at each individually.
    \end{enumerate}
    
\section{Background}
Being able to infer if an individual's data was used for training has a variety of possibly severe implications with regard to privacy violations, generically leading to the discovery -- via conflation with other public information -- of private information on the individual, or information on the medical data sourcer, or both. 
For individuals, leaking knowledge about membership may cause an attacker to realize that a relationship between the individual and the healthcare entity exists, which is a problem as the attacker may have collected additional metadata about the individual that could be a focus for further attacks. Membership information could also be part of a linkage attack, as both the image and whether it was used for training may imply additional private information about the individual. Additionally, the individual may not want the relationship itself to be known, and violating that desire may erode the trust placed in the healthcare entity. Finally, membership inference attacks can be considered gateways to even more critical attacks termed {\em training data reconstruction attacks} which allow for recovery of training data from trained models \cite{carlini2020extracting}.

As the classifier is the main mode of leakage, most defenses~\cite{DP,memguard,mixup,NDSS,globalloss}\yinzhi{Add citations?} focus on augmenting the model either during training or inference.  These schemes all rely on the modeler to actually implement these measures on the given data, and the medical data sourcer trusting the modeler or attempting to audit the classifier directly. However, the incentives for the modeler typically favor pure performance over other concerns, potentially causing privacy to be an afterthought. Most ways to influence these incentives are to enforce compliance through regulation such as Health Insurance Portability and Accountability Act (HIPAA), rather than explicitly rewarding privacy aware models. Moreover, unlike more traditional settings in security, typically the only means for the medical data sourcer to audit the trained model is by the same attacks an attacker would use, either emboldening modelers who maliciously ignore regulation to improve performance, as they could pass a different model from what is actually used to auditors, or not catching ignorant modelers who use bad practices in training. 
\section{Prior Work}

For defending implementations of classifiers against membership inference attacks, there exists methods with both empirical and theoretical successes. For empirical methods, most existing work has focused on regularization of the model during training or inference time. One approach is to reduce overfitting on the training set, as the disparity between training and testing data could arguably be the primary source for identifying training data points. Techniques such as dropout~\cite{globalloss}, which randomly drops part of an activation layer during training, L2 Regularization~\cite{NDSS}, which encourages the smaller weights to help prevent overfitting, and MMD+Mixup~\cite{mixup}. MMD+Mixup combines Mixup, a data augmentation technique that samples random linear interpolations between two data points for both the image and the label, and an MMD regularization term that tries to match the average probability vector between training and validation data. 

There are also methods that use an surrogate adversary to defend against membership inference attacks. For example, Nasr et al.~\cite{previous_work3} turns training into a minimax procedure, where the classifier is trying to both classify the training data correctly while fooling an adversary, which in turn is trained to distinguish between training and reference data. Another adversarial method called MemGuard~\cite{memguard} is notable, as it only affects inference of the trained model and is utility preserving. It trains a surrogate to determine membership using the trained model's logits as input like~\cite{previous_work3}, and minimally perturbs the trained model's logits to fool the surrogate while keeping the same argmax or predicted label. Finally, there are methods that use Generative Adversarial Networks (GANs) to create synthetic data that does not contain identifying information \cite{joshi_2019}, where we focus on curating the generated imagery directly. However, trusting external collaborators with training these models without evaluating the samples can be perilous, as these collaborators could exfiltrate the private data out using the GAN as \cite{liu2020subverting} shows. For theoretical methods, the most prominent example is differential privacy (DP), notably DP-Adam~\cite{DP}, which primarily adds noise to gradients to eliminate that minor perturbations in the gradient that could leak the identity of the data point.

Although most methods that are used to address privacy demonstrate empirical successes, only a subset of these methods offer theoretical guarantees for their performance. Most methods with no guarantees, notably dropout, L2 regularization, and Mixup, were lifted from work studying generalization. Adversarial methods in the former category typically use a specific adversary to defend against attacks, and a stronger adversary may nullify such defenses. Turning training into an adversarial game as with \cite{previous_work3} also increases the complexity of training classifiers. For methods with theoretical backing, using differential privacy is, in general beyond a notion of a privacy budget, opaque to the modeler, causing issues such as inducing disparity with respect to subpopulations \cite{bagdasaryan2019differential}.

\begin{figure}[t!]
\centering 
\includegraphics[width=\linewidth]{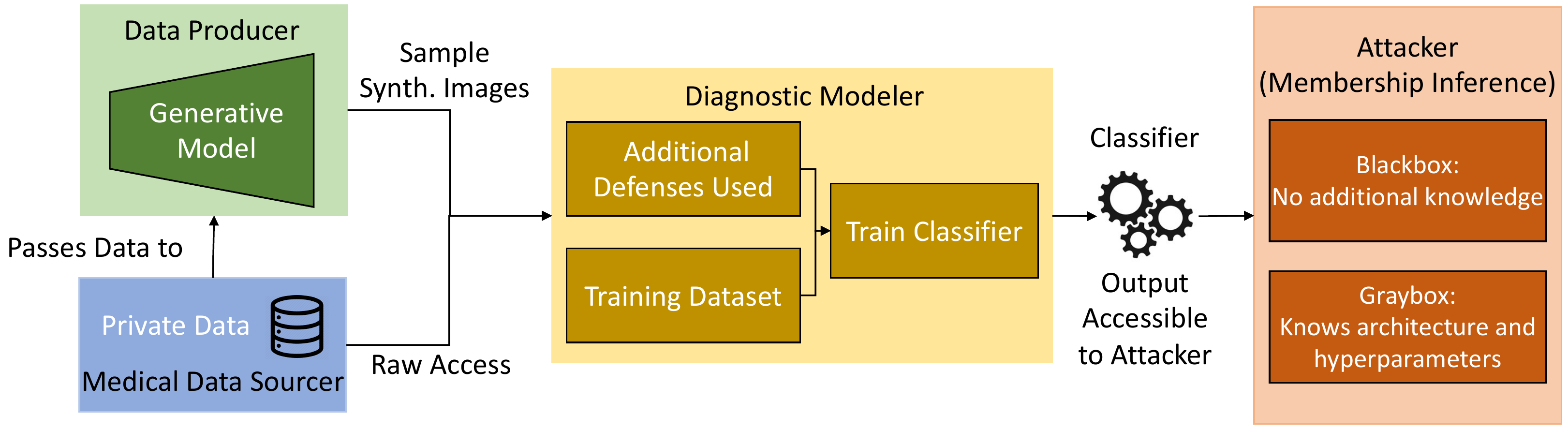}
     \vspace{-0.25in}
    \caption{Overview of our proposed approach and division of duties among agents.} \label{overview}\vspace{-0.25in}
\end{figure}

\section{Methodology}
First, we describe the roles the medical data sourcer, the data producer, the diagnostic modeler, and the attacker play in our methodology, with an overview given in Figure \ref{overview}.

The medical data sourcer is a healthcare entity, such as a hospital, that both has amassed a collection of valuable and private data from individuals and is bound by an institutional review board or other privacy regulator. Alongside ensuring data privacy compliance, the regulator also determines how two images are considered to have the same identity, an equivalence relation $\sim_I$ operating on biometrics such as blood vessels in the retina, and how an image $x$ should be de-identified to produce $\Hat{x}$ such that $x\ \cancel{\sim}_I \Hat{x}$. 

The data sourcer is approached by the diagnostic modeler for access to the private data, needed to learn a diagnostic classifier for some medical task, such as diagnosing diabetic retinopathy. At the end of their process, the modeler desires to have a model that generalizes the task to unseen data, and is accessible in some way by model consumers and attackers. Although the modeler simply needs the private data to create a classifier that works on novel data, common schemes to de-identify typically harm generalization without consideration by the modeler, as novel images are unlikely to have de-identifying alterations like masks. Consequently, most modelers are instead authorized to have raw access to the private data, complicating the liability of the medical data sourcer.

In the case where the data sourcer wants to mitigate privacy leaks from sharing too much data, they turn to a data producer to de-identify images, as they might not have the technical expertise themselves. More traditional methods of doing this primarily involve censoring areas of the image to hide identifiable information as well as hiding metadata. However, this does not preserve the realism of the images, harming performance of the resulting model. Consequently, the data producer can instead use other methods, such as generative adversarial networks, to defend against attackers of the final model attempting to determine the membership of data points, both detailed next.

\subsection{Threat Model}

We define our threat model for classifiers using similar notation as in~\cite{carlini2020attack}:

\begin{definition}[Membership Inference Attack] \hspace{0.03in}
 $D = \{(x_1, y_1),  \dots, (x_n, y_n)\}$ denotes a dataset, where $x_i$ being images and $y_i$ the labels, sampled from $p(x,y)$, acquired by the modeler from either the data producer or the data sourcer. This dataset, not known by the attacker, is used to train a classifier $F$ using training settings $S$, i.e. hyperparameters, optimizers, and architecture used but not any defenses used. Given access to $p(x,y)$, the goal of the attacker
is to choose $(x^*,y^*)$ in the support of $p(x,y)$ such that $x^* \sim_I x_i,$ for an equivalence relation $\sim_I$ and any $i \in [n]$, under one of two settings:

\begin{itemize}
\item \textbf{\em Blackbox setting:} where the attacker has access to the model as an oracle $G(x)$, i.e. whereby the attacker is able to take an input datum  and get the probabilities over $y$ produced by $F(x)$.
\item \textbf{\em Graybox setting:} which allows the adversary access to both $G(x)$ and $S$.
\end{itemize}
\end{definition}

\noindent and the following assumption:

\begin{assumption}[Locality of Inference]
If a data point x is used to train a classifier F, then for any data point $\Tilde{x}\neq x$, $x\ \cancel{\sim}_I \Tilde{x}$ implies that it is not possible for an attacker to find $x^*$ using any level of access to F such that $\Tilde{x} \sim_I x^*$. In other words, a data point used for training $F$ does not leak data points that have different identities.
\label{assumption}
\end{assumption}

As we are explicitly not focused on reconstruction attacks where the attacker could learn $p(x, y)$ directly, the attacker is assumed to have ground truth knowledge of it. $\sim_I$ can be interpreted as comparing the identity of the two images, and an equivalence means that they have the same identity.

\subsection{Approach for Data Producer to Defend Privacy}
In order to produce data points that are usable by the modeler, there are three desirable properties for the generated data: (a.) to preserve the original task, i.e. for classification this means being aligned with a certain class, (b.) to be realistic so that the classifier trained on this dataset can generalize, and (c.) to ensure that the generated data is not equivalent to the original data in the sense of $\sim_I$, which by Assumption \ref{assumption} is sufficient to preserve privacy. (a.) and (b.) can be effectively resolved by resampling from $p(x,y)$, the true data distribution. However, the original way to sample from this distribution is to acquire data from individuals, where privacy concerns arise. Consequently, the data producer wants to construct a surrogate distribution $\Tilde{p}(x,y)$ from the private data that should mimic the true distribution, but without further interaction with any individuals. 

\begin{figure}[t!]
\begin{subfigure}{0.49\textwidth}
\centering 
\includegraphics[width=\linewidth]{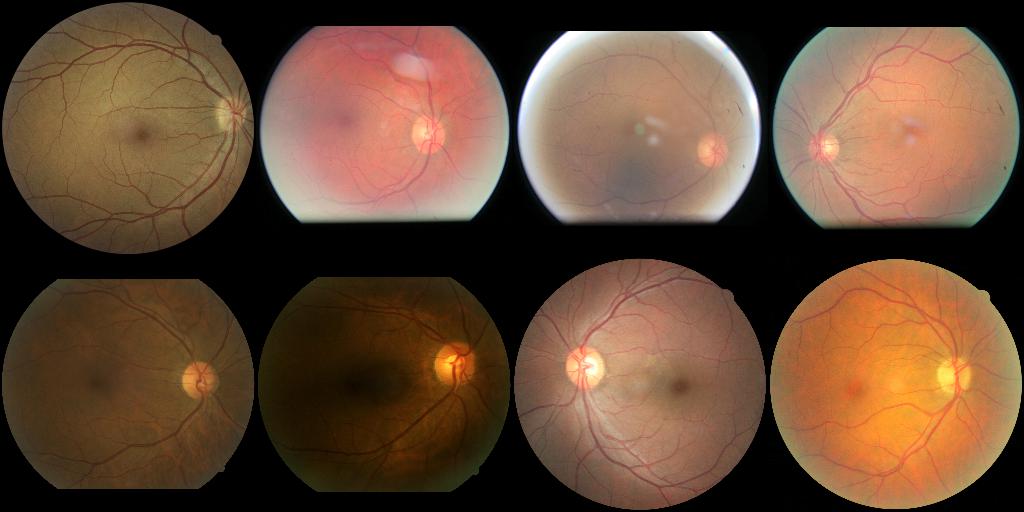}
     \vspace{-0.2in}
    \caption{Healthy}
\end{subfigure}
    \begin{subfigure}{0.49\textwidth}
\centering 
\includegraphics[width=\linewidth]{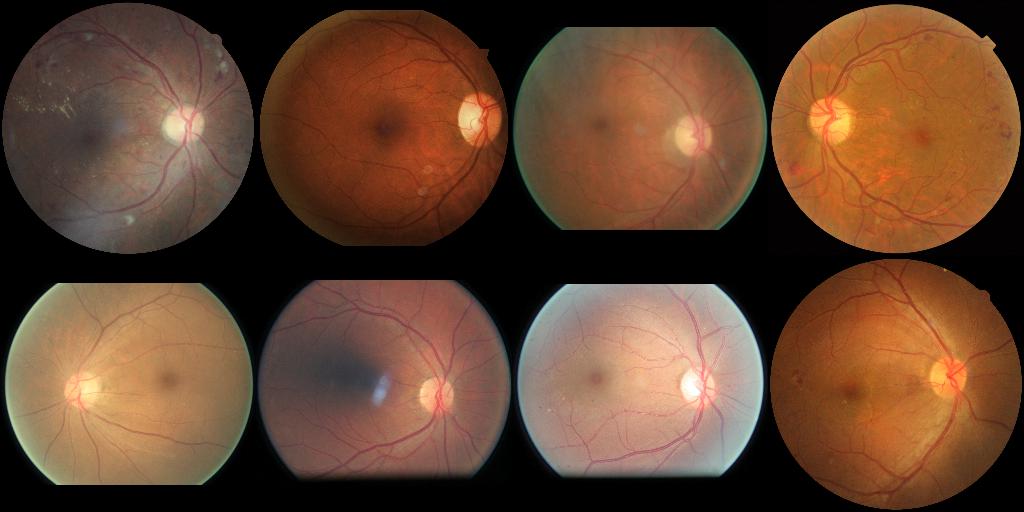}
    \vspace{-0.2in}
    \caption{Diseased}
\end{subfigure}
\caption{Example Real Retinal Images. Labels denote the severity of the diabetic retinopathy from 0 to 4, where 0,1 are taken to be healthy and 2,3,4 to be diseased.}
\label{real_data}
\end{figure}

\begin{figure}[tb!]
\centering 
 \includegraphics[width=\linewidth]{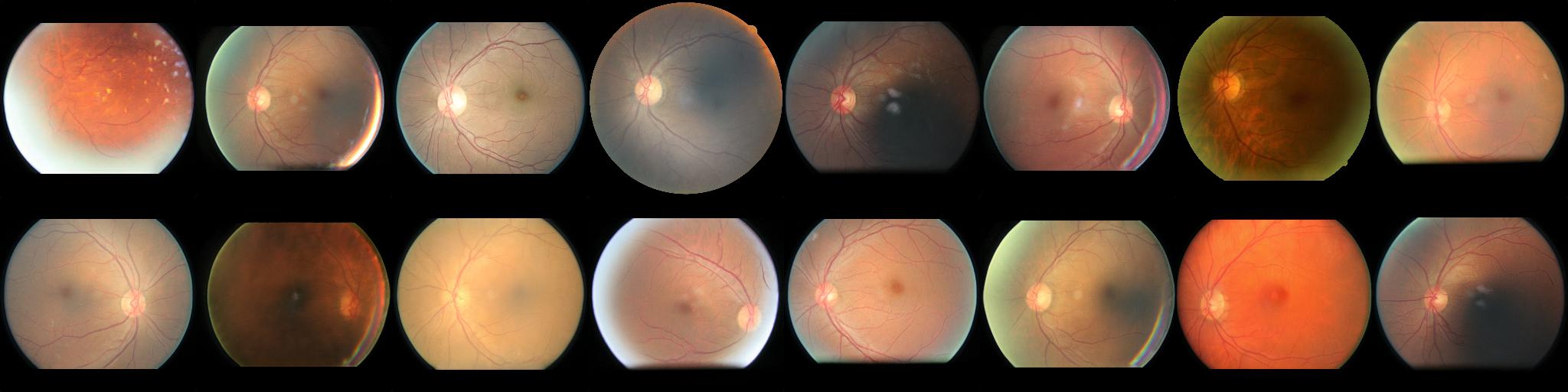}
    \vspace{-0.2in}
    \caption{Example Synthetic Retinal Images from Data Producer.} \label{fig:synthetic} \vspace{-0.2in}
\end{figure}

For this work, we focus on the data producer using Generative Adversarial Networks (GANs)~\cite{goodfellow2014generative}, namely StyleGAN2-ADA~\cite{karras2020training}, to generate this synthetic data. The most desirable properties of GANs are their ability to create realistic data that should conform to the true data, which we leverage here. ~\cite{karras2020training} also includes provisions for training on smaller datasets, enhancing its usefulness on medical imagery. To model $p(x,y)$, the generator and discriminator are made conditional on the label $y$. The data producer can then fix the desired label and sample from the generator to create the data to pass to the modeler. See Figures \ref{real_data} and \ref{fig:synthetic} for real and synthetic data examples. 

To fully satisfy (c.), influencing the sampling to move away from the original points is done by rejecting samples that are equivalent with respect to $\sim_I$. However, it is difficult to fully specify $\sim_I$ in a mathematical form, beyond simple, incomplete measures such as those based on a threshold on the $L^2$ distance in the raw pixel space, which we use in this work. In real world settings, $\sim_I$ will need to be carefully defined using known indicators of identity, i.e., blood vasculature, in order to ensure patient privacy. Consequently, samples are ensured not to immediately return members of the training dataset, and we argue there exist realistic samples that are not identifiable as our generator is continuous. Namely, even if most of the probability support is on the original dataset, there exist interpolations between these points that are sampled and made realistic by GAN training. 

 
 \yinzhi{I deleted the comments on InstaHide. It reads too negative. If we want to mention it, we can say it in the threat model that the training dataset is private to the attacker. }

\subsection{Novel Metric Balancing Utility and Privacy}
As the medical data sourcer determines what degree of access the modeler has to the original data, metrics that combine the utility and the privacy leakage of the final model are needed for determining the level of access.
Indeed, we posit that the field of security and privacy would benefit from the design of more effective metrics that capture possible tradeoffs between utility and privacy:  encouraging higher accuracy for accomplishing the task, i.e. utility, and attenuating accuracy of the attackers in breaching privacy of the diagnostics model, i.e. privacy. We propose here to use a novel metric modeled after the popular F1 score, and which would measure the harmonic mean of the classifier's accuracy and the attack's error rate:
\begin{equation} \footnotesize
    P1(D)_{Attack} = 2*\frac{(Acc_{Task, D}) * ( 1 - Acc_{Attack,D})}{(Acc_{Task,D})  + (1 - Acc_{Attack, D})}
\end{equation}

\noindent where $Acc_{Task, D}$ denotes the accuracy of the defended model, 
and  $Acc_{Attack,D}$ denotes the accuracy of the attacker on the defended model.

\yinzhi{What does 0, 1, 2, 3, 4 mean in Labels of Fig. 2? Can we explain them in the caption?}

\section{Experiments}
As this work is focused on the perspective of the medical data sourcer worried about privacy violations, we focus on the worst case scenario of overfitting the trained classifiers as much as possible \cite{mixup}. To mimic the varying levels of data access the data sourcer can provide to control privacy leaks, we combine the synthetic data and real data for the modeler's training set at different proportions while keeping the overall size constant. 

We use ResNet50, initialized to pretrained ImageNet weights, as our diagnostic classifier, and fix the number of epochs trained to 15 for all experiments, well in excess of what achieves the best accuracy and overfitting. The Adam optimizer is used with a batch size of 64, a learning rate of $5*10^{-5}$, with $\beta_1 = 0.9$ and $\beta_2 = 0.999$, and no validation set is used for the baseline model. The training set for the classifier is used to train the GAN, with default training settings besides a batch size of 32, learning rate of $2.5*10^{-3}$, and R1 penalty of 1. For each mixture of synthetic data and real data, we evaluate both training with no defenses and training with MMD+Mixup. We partition the training dataset into $80\%$ training and $20\%$ validation. The interpolation weight for Mixup is sampled from a $beta(\lambda, \lambda)$ where $\lambda$ is between $\{0, 10\}$. $\lambda = 5$ was found to perform the best with regards to our combined metric, and is shown in both tables. On real data only, we also evaluate on MemGuard, with the same settings from the original work. The task and attack accuracy are evaluated with respect to the modeler's datasets.

For the adversary attempting to determine the membership of training images, they use two attacks from \cite{globalloss}, loss-threshold (Loss-Thre) attacks or label-only attacks. Loss threshold attacks involve the attacker training their own model, called a shadow model, on data they themselves have collected. This shadow model should match the original classifier as much as possible, and this model in the Graybox setting is trained using the same architecture (pretrained ResNet50) and hyperparameters as for our diagnostic classifier in the previous paragraph. For the Blackbox setting, as the attacker does not know the true training settings, we use a pretrained VGG16 network instead and change the number of epochs trained to 9. The remaining configuration is the same as for the graybox and diagnostic classifier. 

Once this shadow model is trained, the attacker can use it to determine statistics about the training and test set used to train the shadow model, and assume these statistics also hold for the classifier being attacked. Namely, the attacker determines the average loss (cross-entropy for these experiments) of the shadow model's training set, so then any image having a cross entropy with respect to the classifier smaller than this average  is a member of the classifier's training dataset. The label-only attack is much simpler, and does not require a shadow model be constructed and thus is independent of the attach setting. This attack simply predicts  membership in the classifier's training dataset for a target image if the classifier produces the correct answer for this image, i.e., training set members are more likely to be correct compared to data that was not trained on.


    
    
    


\begin{table*}[!t] \vspace{-0.3in}
\setlength{\tabcolsep}{3pt}
   \renewcommand{\arraystretch}{1.2}
   \scriptsize
    \caption{Task and Attack Accuracies (\%) of various defenses with different percentages of synthetic data. 95\% CIs are in parenthesises, and numbers in bold are the best. 
     } 
    \label{acc_results}
	\centering
	\begin{tabular}{c|c|c|c|cc|cc}
	    \toprule
	    &\multirow{3}{*}{\parbox{1.2cm}{\centering \textbf{Raw Data Access \%}}}&\multirow{3}{*}{\textbf{Defense}} & \multirow{3}{*}{\textbf{$Acc_{Task, D}$}} & \multicolumn{4}{c}{\textbf{$Acc_{Attack, D}$}} \cr
	    \cline{5-8}
	    &&&& \multicolumn{2}{c|}{\textbf{Blackbox}} & \multicolumn{2}{c}{\textbf{Graybox}} \cr
	    \cline{5-8}
        &&&& \textbf{Loss-Thre} & \textbf{Label-Only} &  \textbf{Loss-Thre} & \textbf{Label-Only} \cr
        \midrule
        \midrule
        \multirow{2}{*}{\parbox{1.3cm}{\centering \textbf{Synthetic Data Only}}}&\multirow{2}{*}{\textbf{0\%}}&\textbf{No Defense} & 68.77 (0.91)&49.84 (0.69)&49.90 (0.69)&\textbf{49.85 (0.69)}&49.95 (0.69) \cr
        &&\textbf{MMD+Mixup}& 73.30 (0.87)&\textbf{49.67 (0.69)}&\textbf{49.79 (0.69)}&49.99 (0.69)&\textbf{49.79 (0.69)} \cr \cline{1-8}\cline{1-8}
        \multirow{6}{*}{\parbox{1.4cm}{\centering \textbf{Synthetic/ Real Mixture}}}&\multirow{2}{*}{\textbf{25\%}}&\textbf{No Defense} & 73.54 (0.86)&53.35 (0.69)&53.10 (0.69)&54.98 (0.69)&53.10 (0.69) \cr
        &&\textbf{MMD+Mixup}& 74.10 (0.86)&52.65 (0.69)&52.17 (0.69)&50.05 (0.69)&52.17 (0.69) \cr \cline{2-8}
        &\multirow{2}{*}{\textbf{50\%}}&\textbf{No Defense} & 72.95 (0.87)&57.49 (0.69)&56.70 (0.69)&60.31 (0.68)&56.70 (0.69) \cr
        &&\textbf{MMD+Mixup}& 73.80 (0.86)&51.91 (0.69)&52.86 (0.69)&50.08 (0.69)&52.86 (0.69) \cr \cline{2-8}
        &\multirow{2}{*}{\textbf{75\%}}&\textbf{No Defense} & 75.14 (0.85)&60.43 (0.68)&59.21 (0.68)&66.03 (0.66)&59.21 (0.68) \cr
        
        &&\textbf{MMD+Mixup}& 74.75 (0.85)&54.52 (0.69)&55.41 (0.69)&50.01 (0.69)&55.41 (0.69) \cr \cline{1-8}\cline{1-8}
        \multirow{3}{*}{\parbox{1.4cm}{\centering \textbf{Real Data} \textbf{Only}}}&\multirow{3}{*}{\textbf{100\%}}&\makecell{\textbf{No Defense} \\\textbf{(Baseline)}} & 73.24 (0.87)&64.25 (0.66)&62.70 (0.67)&66.64 (0.65)&62.70 (0.67) \cr 
        &&\textbf{MMD+Mixup} &\textbf{75.52 (0.84)} &61.80 (0.67)&59.23 (0.68)&50.08 (0.69)&59.23 (0.68) \cr
        &&\textbf{Memguard} &73.24 (0.87)&63.87 (0.67)&62.70 (0.67)&63.87 (0.67)&62.70 (0.67) \cr
        \bottomrule
	\end{tabular} 
	\vspace{-0.5cm}
\end{table*} 

\subsection{Dataset}
For our experiments, we use the EyePACs dataset  from Kaggle \cite{KaggleEyePACSdataset}, originally used for a Diabetic Retinopathy Detection challenge. The dataset includes 88,703 high-resolution retina images taken under a variety of imaging conditions and each image has a label ranging from 0 to 4,  representing the presence and severity of diabetic retinopathy. 

We select 10,000 random images each for modeler's training and testing set and the attacker's training and testing set, all disjoint. The images are cropped to the boundary of the fundus, and resized to $256$ by $256$ pixels. The GAN is trained on this data, and the classifier has additional processing for contrast normalization. The GAN used is Stylegan2-ADA \cite{karras2020training} using Adaptive Data Augmentation with a minibatch of 32, minibatch standard deviation layer with 8 samples, a feature map multiplier of 1, learning rate of 0.0025, gamma of 1, and 8 layers in the mapping network. The rest of the configuration remains the same from the original work.

To reject samples that are identifiable, the threshold we use is the minimum $L^2$ distance in pixel space between images in the training dataset, as we know that images in the training dataset are distinct. Consequently, a synthetic image that is a distance smaller than this threshold away from any training image is taken to have the same identity and rejected.  The actual threshold computed was 2328.96 for the dataset we use, where the minimum pixel value is 0 and and the maximum 255 for computing the threshold. Consequently, none of the synthetic images we generated are within this threshold to a point, and, as the threshold is greater than zero, not equivalent to any training image.

When evaluating the quality of the generated images, we use the Frechet Inception Distance (FID) \cite{heusel2018gans}, a standard metric in GAN literature that computes the distance between real image representations and synthetic image representations taken from a layer of the Inception network. More specifically, each set of representations is assumed to be Gaussian, so the Frechet distance between the two distributions can be computed. The FID of the generator we used for this work is 4.12. In addition, we asked three highly trained retinal specialists to critically assess  images in Figure \ref{fig:synthetic} for vessels' realism, with two questions: how many look real? (answers: 16, 12, and 7) and how many have potential vasculature issues? (answers: 1, 4, 9).  While empirical results show the potential of our method for classification, these clinical assessments suggest future work incorporating a loss term to regularize the structure of the resulting blood vessel branching and to increase their realism which would make the generated images useful for example as private datasets that are also adequate for resident training. From our qualitative evaluation, we believe that our generator produces images of an acceptable quality for training, and assign these images the labels that were used to generate them (as our generator is conditional on the class label).

\subsection{Results} \yinzhi{Add references to the tables and figures in the text? (like Table 1, Table 2, Figure 1, Figure 2)}
In Table \ref{acc_results}, we see that using only synthetic data for training completely defeats the attacker (attack probability is close to chance), at the cost of decreased accuracy. Using MMD+Mixup on synthetic data only improves the task accuracy to be on par with the original classifier, and  close to MMD+Mixup with only real data. Due to its utility preserving nature, MemGuard does not outperform MMD+Mixup for attack accuracy nor task accuracy. For most attacks, the attack accuracy with no defenses increases roughly linearly with the amount of real data in the training dataset. With the MMD+Mixup defense, the relationship is less clear, possibly due to the dataset partitioning. For the combined metric $P1(D)_{Attack}$ in Table \ref{p1_results}, we see that synthetic data only with MMD+Mixup is the best of breed against all attacks except the Graybox loss-threshold attack, for which the best performance is obtained by the MMD+Mixup on real data only due to its higher task accuracy, but the difference between that and the second best method which uses 100\% synthetic data with MMD+Mixup is barely significant.

\begin{table*}[!t] \vspace{-0.2in}
\setlength{\tabcolsep}{3pt}
   \renewcommand{\arraystretch}{1.2}
   \scriptsize
    \caption{$P1(D)_{Attack}$ $\in [0,100]$ where larger is better for various settings and defenses.} 
    \label{p1_results}
	\centering
	\begin{tabular}{c|c|c|cc|cc}
	    \toprule
	    &\multirow{3}{*}{\parbox{1.2cm}{\centering \textbf{Raw Data Access \%}}}&\multirow{3}{*}{\textbf{Defense}} &  \multicolumn{4}{c}{\textbf{$P1(D)_{Attack}$}} \cr
	    \cline{4-7}
	    &&& \multicolumn{2}{c|}{\textbf{Blackbox}} & \multicolumn{2}{c}{\textbf{Graybox}} \cr
	    \cline{4-7}
        &&& \textbf{Loss-Thre} & \textbf{Label-Only} &  \textbf{Loss-Thre} & \textbf{Label-Only} \cr
        \midrule
        \midrule
        \multirow{2}{*}{\parbox{1.3cm}{\centering \textbf{Synthetic Data Only}}}&\multirow{2}{*}{\textbf{0\%}}&\textbf{No Defense} &58.01 (0.63)&57.97 (0.63)&58.00 (0.63)&57.94 (0.63) \cr
        &&\textbf{MMD+Mixup} &\textbf{59.68 (0.62)}&\textbf{59.60 (0.62)}&59.46 (0.62)&\textbf{59.60 (0.62)} \cr \cline{1-7}\cline{1-7}
        \multirow{6}{*}{\parbox{1.4cm}{\centering \textbf{Synthetic/ Real Mixture}}}&\multirow{2}{*}{\textbf{25\%}}&\textbf{No Defense} &57.09 (0.61)&57.27 (0.61)&55.85 (0.61)&57.27 (0.61) \cr
        &&\textbf{MMD+Mixup} &57.78 (0.61)&58.14 (0.61)&59.67 (0.61)&58.14 (0.61) \cr \cline{2-7}
        &\multirow{2}{*}{\textbf{50\%}}&\textbf{No Defense} &53.72 (0.61)&54.34 (0.61)&51.41 (0.60)&54.34 (0.61) \cr
        &&\textbf{MMD+Mixup} &58.23 (0.61)&57.53 (0.61)&59.56 (0.61)&57.53 (0.61) \cr \cline{2-7}
        &\multirow{2}{*}{\textbf{75\%}}&\textbf{No Defense} &51.84 (0.60)&52.88 (0.60)&46.79 (0.58)&52.88 (0.60) \cr
        
        &&\textbf{MMD+Mixup} &56.55 (0.61)&55.86 (0.61)&59.91 (0.61)&55.86 (0.61) \cr \cline{1-7}\cline{1-7}
        \multirow{3}{*}{\parbox{1.4cm}{\centering \textbf{Real Data} \textbf{Only}}}&\multirow{3}{*}{\textbf{100\%}}&\makecell{\textbf{No Defense} \\\textbf{(Baseline)}} &48.05 (0.59)&49.43 (0.60)&45.84 (0.59)&49.43 (0.60) \cr 
        &&\textbf{MMD+Mixup}&50.74 (0.59)&52.95 (0.60)&\textbf{60.11 (0.61)}&52.95 (0.60) \cr
        &&\textbf{Memguard}&48.39 (0.59)&49.43 (0.60)&48.39 (0.59)&49.43 (0.60) \cr
        \bottomrule
	\end{tabular} 
	\vspace{-0.5cm}
\end{table*} 
\section{Discussion and Limitations}
Though we have demonstrated that our synthetic data can be used to recover similar accuracies as training with real data in this case, in general GANs have issues with being able to fully represent all modalities of the training dataset. Improvements in generative modeling are still being made that should address this deficiency, but in the meantime, we note that the data sourcer can give raw access to data necessary for the classifier to perform well. We see from Table \ref{acc_results} that the attack accuracy for classifiers with no defenses increases linearly with the amount of real data included, suggesting that each individual data point trained on only violates the privacy of itself  and assumption \ref{assumption} holds. Consequently, the sourcer can still protect most of the population while giving the modeler access to data that can be instrumental to model performance.

Our introduced $P1(D)_{Attack}$ metric uses the overall accuracy as the primary measure of utility, but other measures such as F1 score, precision, or area under the receiver operating characteristic curve (AUC) could be used instead. The accuracy in the formulation of the metric can be substituted by these measures directly, but one concern that becomes more apparent is how to normalize changes in each individual metric such that the actual numerical value matches user preferences. For example, the user may prefer a 3\% increase in accuracy or a 0.05 increase in the F1 score over a 7\% decrease in the attack accuracy, but the current $P1$ formulation uses a fixed, equal weighting between utility and privacy. A simple way to address this is to allow the user to set this weighting, but more investigation will be needed for how well this can match preferences.

Although the equivalence relation $\sim_I$ we use for this work was effective in defeating adversaries, more sophisticated measures can help situations where an image is far away in pixel space but still retains the same identity, such as when artifacts are present or the overall brightness is different. For retinal imagery, these can be based on biometrics such as the retinal vasculature, which can be extracted via vessel segmentation techniques \cite{Li_2020_WACV} and compared between images. For general medical imagery, more robust measures of perceptual similarity can be used, such as structural similarity index measure (SSIM), or metrics such as LPIPS \cite{zhang2018perceptual} that use the intermediate layers of deep networks.

Finally, a limitation of our work is not comparing directly to differential privacy (DP), which should be done in future work. Additionally, assumption \ref{assumption} is reliant on the definition of $\sim_I$ to assume privacy, and, as discussed in the previous paragraph, there could be uncertainty about whether $\sim_I$ captures the true identity. One interesting direction to extend this work is to treat sampling and rejecting synthetic data as a random mechanism in the same way DP treats computing the gradient as a random mechanism, and use DP's notion of a privacy budget in place of the assumption. 
\section{Conclusion}
We propose a novel approach using generative methods to defend membership inference attacks of retinal diagnostics. 
Our evaluation shows that, used alone or in combination of SOTA defenses, it confers significantly reduction in  attack accuracy while minimally impacting the model's worst case utility. 
Our approach can also be improved in the future via better control over the generator and specification of $\sim_I$ to better satisfy privacy advocates, regulators, and modelers together.

\section*{Acknowledgments}
We thank Drs. Bressler, Liu (John Hopkins University (JHU) School of Medicine) and Delalibera (Eye Hospital, Brasilia, Brazil) for their help assessing images in Fig. 3. This work was funded by the JHU Institute for Assured Autonomy.

%
%
%
\bibliographystyle{splncs04}

\bibliography{example_paper}

\end{document}